# Massive suppression of proximity pairing in topological $(Bi_{1-x}Sb_x)_2Te_3$ films on niobium

Joseph A. Hlevyack,[1,2,†] Sahand Najafzadeh,[3,†] Meng-Kai Lin,[1,2] Takahiro Hashimoto,[3] Tsubaki Nagashima,[3] Akihiro Tsuzuki,[3] Akiko Fukushima,[3] Cédric Bareille,[3] Yang Bai,[1,2] Peng Chen,[1,2,4,5] Ro-Ya Liu,[1,2,6] Yao Li,[1,2] David Flötotto,[7] José Avila,[8] James N. Eckstein,[1,2] Shik Shin,[9] Kozo Okazaki,[3*] and T.-C. Chiang[1,2*]

[1]Department of Physics, University of Illinois at Urbana-Champaign, Urbana, Illinois 61801, USA.

[2]Frederick Seitz Materials Research Laboratory, University of Illinois at Urbana-Champaign, Urbana, Illinois 61801, USA.

[3]Institute for Solid State Physics, The University of Tokyo, Kashiwa, Chiba 277-8581, Japan.

[4]Shanghai Center for Complex Physics, School of Physics and Astronomy, Shanghai Jiao Tong University, Shanghai 200240, China.

[5]Key Laboratory of Artificial Structures and Quantum Control (Ministry of Education), Shenyang National Laboratory for Materials Science, School of Physics and Astronomy, Shanghai Jiao Tong University, Shanghai 200240, China.

[6]Advanced Light Source, Lawrence Berkeley National Laboratory, Berkeley, California 94720, USA.

[7]Center for Soft Nanoscience, University of Münster, 48149 Münster, Germany.

[8]Synchrotron SOLEIL and Université Paris-Saclay, L'Orme des Merisiers, BP48, 91190 Saint-Aubin, France.

[9]Office of University Professor, The University of Tokyo, Kashiwa, Chiba 277-8581, Japan.

*Corresponding authors:

T.-C. Chiang, Email: tcchiang@illinois.edu

Kozo Okazaki, Email: okazaki@issp.u-tokyo.ac.jp

Additional author notes:

†Equal contribution

Interfacing bulk conducting topological $Bi_2Se_3$ films with *s*-wave superconductors initiates strong superconducting order in the nontrivial surface states. However, bulk insulating topological $(Bi_{1-x}Sb_x)_2Te_3$ films on bulk Nb instead exhibit a giant attenuation of surface superconductivity, even for films only two-layers thick. This massive suppression of proximity pairing is evidenced by ultrahigh-resolution band mappings and by contrasting quantified superconducting gaps with those of heavily *n*-doped topological $Bi_2Se_3$/Nb. The results underscore the limitations of using superconducting proximity effects to realize topological superconductivity in nearly intrinsic systems.

Topological superconductors are among the most unconventional states of matter, wherein exotic *p*-wave-like pairing and time-reversal symmetry can foster supersymmetry and zero-energy excitations called Majorana bound states [1–7]. These emergent quasiparticles obey a non-Abelian statistics pertinent to topological quantum computing [5–7]. Despite evidence in $Cu_xBi_2Se_3$, the ruthenates, iron-based superconductors, and perhaps K-doped β-$PdBi_2$ [5–10], robust topological superconductor states remain elusive due to the often small superconducting gap and low transition temperature in candidate materials [6–9]. One promising alternative for realizing this state is to couple topological insulators (TIs) to a superconductor (SC) so that *p*-wave-like pairing is induced into the topological boundary states via the superconducting proximity effect [4–7]. Though proximity-induced superconductivity has been confirmed in the TIs $Bi_2Te_3$ and $Bi_2Se_3$ by transport [11–13], scanning tunneling spectroscopy [14–17], and the momentum-resolved probe angle-resolved photoemission spectroscopy (ARPES) [18,19], the requirements for inducing superconductivity into the surface states are mostly unknown. Prior ARPES studies have shown proximity pairing in the nontrivial surface states of $Bi_2Se_3$ prepared on $NbSe_2$ and Nb [18,19], but for $Bi_2Se_3$ interfaced with the *d*-wave SC $Bi_2Sr_2CaCu_2O_{8+\delta}$, the proximity effect is suppressed [20,21]. Various reasons for this contrast are conjectured, including Fermi surface and/or lattice mismatch at the TI/SC interface and the short superconducting coherence length of a *d*-wave SC [20,21]. Furthermore, quantum-mechanical coupling between bulk and surface states likely drives the proximity effect in TIs/Nb, as both states exhibit the same superconducting gap in $Bi_2Se_3$/Nb despite different characters of the Cooper pairs [19]. Moreover, all TI/SCs containing $Bi_2Te_3$ and $Bi_2Se_3$ suffer from intrinsic bulk carrier doping of the TI, complicating any superconductivity signatures in the surface states [14–21], and do not contain TIs in the real sense with conducting boundary states but insulating bulk.

Here, we present ultrahigh-resolution ARPES mappings of clean, bulk insulating $(Bi_{1-x}Sb_x)_2Te_3$ ($x$ = 0.62) on superconducting Nb substrate as a function of TI film thickness and temperature, which together with $Bi_2Se_3$/Nb reveal the pivotal role of bulk states in transiting superconductivity to the surfaces of TIs/Nb. To circumvent the inherent difficulties in growing TIs on Nb substrates [19], we employ a novel cleavage-based flip-chip method, which upon cleaving yields for the first time nearly intrinsic TI films of specific thicknesses, even in the ultrathin-film limit, on superconducting bulk Nb films. Proximity-induced superconducting gaps for slightly $n$-doped $(Bi_{1-x}Sb_x)_2Te_3$/Nb are quantified and strongly contrasted with those of heavily $n$-doped $Bi_2Se_3$/Nb from our prior work, which revealed a substantial proximity-induced gap with a long decay length [19]. Our results demonstrate a massive attenuation of surface superconductivity in $(Bi_{1-x}Sb_x)_2Te_3$/Nb compared to $Bi_2Se_3$/Nb, suggesting that quantum-mechanical coupling between bulk and surface states largely dominates the proximity effect in TIs/Nb.

Prior to growths of $(Bi_{1-x}Sb_x)_2Te_3$ films by molecular beam epitaxy (MBE) at the University of Illinois, 6H-SiC(0001) substrates were repeatedly flash annealed to ~1300 °C to fabricate well-ordered bilayer-graphene surfaces [22]. High purity Bi, Sb, and Te were co-evaporated from an electron-beam evaporator (Bi) and effusion cells (Sb and Te) onto a substrate held at 280 °C. The film growth rate was 1 quintuple layer (QL) (1 QL ≈ 1 nm) every ~12 min.; the alloy ratio $x$ = 0.62 was calibrated *in situ* with a quartz crystal microbalance monitor so that the Fermi level crosses only the surface states just above the Dirac point [23]. Each $(Bi_{1-x}Sb_x)_2Te_3$ sample was post-annealed at 310 °C for ~30 min. to smoothen the film further. Sample quality was confirmed *in situ* with reflection high-energy electron diffraction (RHEED) and ARPES. Ultrathin $Bi_2Se_3$ films were prepared on *in situ* ozone-cleaned $Al_2O_3$(0001) substrates via a "two-step" approach [19]. Bi and Se were co-evaporated onto a substrate held at 220 °C for the first two QL and then at 280 °C for the remaining QL; each sample was post-annealed under a lower Se flux for ~3 hours at 280 °C. To prepare TIs/Nb, a 600 Å-thick polycrystalline Nb film was sputter-deposited onto each TI film at ~25 °C under an Ar gas pressure of 3 mTorr and at a power of 120 W. These samples were flipped over and glued onto a polished copper sheet with low-temperature-curing Ag epoxy. A cleave pin was lastly fixed to the substrate's backside using Torr seal or Ag epoxy. Ultrahigh-resolution ARPES measurements were performed at the Institute for Solid State Physics (ISSP) at The University of Tokyo in a laser-based ARPES setup, which consisted of a vacuum ultraviolet laser (6.994-eV photons), a Scienta HR8000 hemispherical analyzer, and a sample manipulator cooled with superfluid liquid helium. The sample temperature was varied from 15 K to 1.5 K, and the energy resolution was better than 1.5 meV. For details of the Fermi-level calibration with a gold reference, see the Supplemental Material [24] and methods in Refs. [25,26]. ARPES spectra of as-grown TIs and TI/Nb systems were taken at the University of Illinois at 18 K using a He discharge lamp. Synchrotron results were measured at the ANTARES beamline of Synchrotron SOLEIL at 80 K and an energy resolution of 15 meV.

A schematic of bulk-carrier doping in the band structure of prototypical TIs such as $Bi_2Se_3$ and $Bi_2Te_3$ is shown in Fig. 1(a), where $n+$ doping from chalcogen vacancies pins the Fermi level to the conduction band [23,27]. This carrier doping is a serious issue as the unique transport properties of Dirac fermions are obscured by trivial bulk carriers, thus complicating any signatures of the proposed Majorana bound state in TI/SC systems [21,23]. One method for isolating surface from bulk carriers is to introduce bulk Sb alloying into the parent compound $Bi_2Te_3$ [Fig. 1(b)] [23]. Both the Dirac point and surface states can be isolated from the conduction and valence bands by tuning the alloy ratio $x$ in $(Bi_{1-x}Sb_x)_2Te_3$ [23,28–30]. Slightly $n$-doped ultrathin $(Bi_{1-x}Sb_x)_2Te_3$ films ($x = 0.62$) of thicknesses $N$ = 2–10 QL are grown on bilayer-graphene-terminated 6H-SiC(0001) (BLG/SiC) substrates. A representative *in situ* RHEED pattern of a 2 QL film is shown in Fig. 1(c); the sharp RHEED pattern is indicative of a well-ordered film. ARPES maps and corresponding second-derivative spectra taken along $\overline{\Gamma K}$ of an as-grown 2 QL film [left and right panels of Fig. 1(d), respectively] show a distinctive but gapped Dirac cone. In the second-derivative map of Fig. 1(d), a small hybridization gap (~0.08 eV) is discernible in the Dirac cone, a quintessential feature in ultrathin TI films arising from quantum tunneling between surface states on both film boundaries [31,32]. No conduction bands are visible in these maps. Thus, these $(Bi_{1-x}Sb_x)_2Te_3$ films are bulk insulating with only the surface states crossing the Fermi level even in the ultrathin-film limit.

Figure 1(e) shows an illustration of our flip-chip method for preparing TIs on the isotropic *s*-wave superconductor Nb. Once $(Bi_{1-x}Sb_x)_2Te_3$/Nb ($x$ = 0.62) films of thicknesses $N$ = 2–10 QL are prepared on BLG/SiC, a 600 Å-thick polycrystalline Nb film is sputter-deposited onto each film at ~25 °C [step 1. in Fig. 1(e)]. These samples are flipped over [step 2. in Fig. 1(e)] and glued onto a polished copper plate using Ag epoxy with the Nb surface facing downwards. Lastly, the sample structure is topped with a cleave pin [step 3. in Fig. 1(e) and Fig. 1(f)]. Each sample is introduced into the ARPES system, and prior to the measurement, the sample is cleaved *in situ* by pushing sideways against the cleave pin, exposing a fresh TI surface [Fig. 1(g)]. Per prior studies on Bi$_2$Se$_3$/Nb [19] and ARPES maps in Figs. 1(h) and 1(i), the cleavage occurs exactly at the substrate/TI interface, which involves incommensurate van der Waals bonding, as the weakest point in this sample structure, resulting in a TI film of a thickness predetermined by the MBE growth on a superconducting Nb substrate.

ARPES band maps and corresponding second-derivative spectra of $(Bi_{1-x}Sb_x)_2Te_3$/Nb taken along $\overline{\Gamma M}$ and with TI film thicknesses $N$ = 2–4 QL [Figs. 1(h) and 1(i), respectively] exhibit sharp Dirac cones. No conduction bands are discernible, demonstrating that all films prepared on Nb are all *n*-type with no bulk-like states at the Fermi level. An apparent closing of the hybridization gap in the 2 QL spectra of Figs. 1(h) and 1(i) can be attributed to interactions at the TI/Nb interface which lift the degeneracy between the topological surface states on the TI film's two boundaries, each having penetration depths of about 1 QL [31,33]. The decoupling of the surface states at the films' two boundaries well isolates those at the top surface from the substrate, even at a film thickness of just 2 QL. These ARPES results establish a novel nearly intrinsic TI/SC system wherein the density of bulk-like states at the Fermi level is fully suppressed even in the ultrathin-film limit, quite different from the TIs $Bi_2Te_3$ and $Bi_2Se_3$ prepared on a SC [14–21].

Ultrahigh-resolution ARPES band maps taken about the zone center are measured at temperatures $T = 1.5$ to 15 K for all samples. Example datasets for 4 QL $Bi_2Se_3$/Nb [19] and 4 QL $(Bi_{1-x}Sb_x)_2Te_3$/Nb are shown in Fig. 2. At $T = 10$ K, these two systems exhibit thermally-broadened Fermi edges [top panels of Figs. 2(b) and 2(f)], but at $T < 2.2$ K, their behaviors diverge remarkably: 4 QL $Bi_2Se_3$/Nb exhibits coherence peaks, leading edge shifts, and the onset of a superconducting gap, while 4 QL $(Bi_{1-x}Sb_x)_2Te_3$/Nb shows only a sharpened Fermi edge with no identifiable coherence peak and no clear leading edge shift [bottom panels of Figs. 2(b) and 2(f), respectively]. This sharp contrast at low temperatures is further highlighted by symmetrizing the momentum-resolved maps about the Fermi level [Figs. 2(c) and 2(g)]. To underscore the divergence between these TI/Nb systems and precisely quantify leading edge shifts, *k*-space integrated energy distribution curves (EDCs) and corresponding symmetrized EDCs taken as a function of temperature are summarized in Fig. 2(d) for 4 QL $Bi_2Se_3$/Nb and Fig. 2(h) for 4 QL $(Bi_{1-x}Sb_x)_2Te_3$/Nb. As the sample is cooled from $T = 10$ K to $T < 2.2$ K, clear leading edge shifts and coherence peaks develop in the EDCs of 4 QL $Bi_2Se_3$/Nb [Fig. 2(d), top panel] but are both lacking in 4 QL $(Bi_{1-x}Sb_x)_2Te_3$/Nb [Fig. 2(h), top panel]. The symmetrized data of 4 QL $Bi_2Se_3$/Nb shows the onset of a superconducting gap [Fig. 2(d), bottom panel], while that of 4 QL $(Bi_{1-x}Sb_x)_2Te_3$/Nb flat lines within the noise/uncertainty level at all temperatures [Fig. 2(h), bottom panel], indicating that no superconducting gap is detected at these temperatures. Thus, while 4 QL $Bi_2Se_3$/Nb unambiguously exhibits proximity-induced superconductivity, 4 QL $(Bi_{1-x}Sb_x)_2Te_3$/Nb shows no signs of superconductivity within the experimental energy resolution.

Due to the massive suppression of proximity-induced superconductivity in 4 QL $(Bi_{1-x}Sb_x)_2Te_3$/Nb, thickness-dependent ARPES spectra as a function of temperature are taken to reveal whether superconductivity can be detected in $(Bi_{1-x}Sb_x)_2Te_3$/Nb at reduced TI film thicknesses. Figure 3 displays the *k*-space integrated EDCs of $(Bi_{1-x}Sb_x)_2Te_3$/Nb as a function of both TI film thickness and temperature compared with $Bi_2Se_3$/Nb [right and left panels of Fig. 3, respectively]; Fig. 4 shows these same EDC datasets symmetrized about the Fermi level. For comparison, *k*-space integrated EDCs and associated symmetrized EDCs of a bare polycrystalline Nb sample are shown at the tops of Figs. 3 and 4, respectively. In all measured $(Bi_{1-x}Sb_x)_2Te_3$/Nb ($N$ = 2–5 QL), the EDCs in Fig. 3 as a function of temperature seemingly only exhibit a sharpening of the Fermi edge upon cooling from $T$ = 10 K and no leading edge shift; all symmetrized EDCs in Fig. 4 thus show no development of a superconducting gap, only occasional bumps and wiggles in the EDCs at low temperatures attributable to slight errors in Fermi-level alignment on the order of the analyzer's energy resolution (~100 $\mu$eV) [25,26]. This is in stark contrast to $Bi_2Se_3$/Nb [19]: Clear leading edge shifts and coherence peaks in the EDC curves in Fig. 3 arise upon cooling from $T$ = 10 K for all TI film thicknesses sampled ($N$ = 4–10 QL); the symmetrized data of $Bi_2Se_3$/Nb in Fig. 4 shows a superconducting gap for all thicknesses at the lowest temperature. Similar statements for the Nb reference are duly noted as with $Bi_2Se_3$/Nb. Hence, despite a clear onset of superconductivity for the Nb reference and all $Bi_2Se_3$/Nb ($N$ = 4–10 QL), no proximity-induced superconductivity can be observed in $(Bi_{1-x}Sb_x)_2Te_3$/Nb ($N$ = 2–5 QL).

The null results for superconductivity in $(Bi_{1-x}Sb_x)_2Te_3$/Nb are highly illuminating in regards to the coupling mechanism that may lead to topologically nontrivial superconducting pairs at surfaces. One possible cause for superconductivity at the surface of TIs/Nb is quantum-mechanical coupling between bulk and surface states [19]. Cooper pairs transferred to the $Bi_2Se_3$ film by Andreev reflection at the TI/SC interface reach the surface via the bulk states and/or quantum tunneling of topological interface states to the surface, though Cooper pairs lose their phase coherence with increasing distance away from the TI/SC interface [34]. At the probed $Bi_2Se_3$ surface, both the topological surface and bulk states exhibit the same superconducting gap, implying that any superconducting properties impressed upon the bulk states are transferred to the surface states [19]. If there are no bulk states at the Fermi level of the TI, as in $(Bi_{1-x}Sb_x)_2Te_3$/Nb ($x$ = 0.62), then there are no bulk states/carriers which can transfer superconductivity to the topological surface states. For selenide/telluride-based TIs, a thickness of 2 QL is also the limit for any semblance to the bulk system, due to quantum tunneling between the topological surface states from both film boundaries [31,32]. However, since this resonant tunneling between the surface states at the two boundaries of the $(Bi_{1-x}Sb_x)_2Te_3$ films is suppressed in our case [Figs. 1(h) and 1(i)], superconducting pairing in the nontrivial boundary states at the top surface can only arise from direct coupling of these surface states to the superconducting pairs from the Nb substrate. Arguably, superconductivity at the surface may be strongly suppressed because of the very short penetration depth, about 1 QL, of the surface states [31]. The current ARPES results of $Bi_2Se_3$/Nb and $(Bi_{1-x}Sb_x)_2Te_3$/Nb together support this picture, suggesting that bulk states are key for transiting superconductivity to the TI surface over any appreciable length scale. Along with employing a flip-chip technique for preparing unique TI/SC systems, our study not only unveils a mechanism for tuning the strength of the proximity effect in TIs/Nb but also underlines the inherent limitations of using the proximity effect to realize topological superconductivity in TI/Nb systems.

This work was supported by the U.S. Department of Energy (DOE), Office of Science (OS), Office of Basic Energy Sciences, Division of Materials Science and Engineering, under Grant No. DE-FG02-07ER46383 (T.C.C.), and by the Ministry of Education, Culture, Sports, Science and Technology, Japan, Japan Society for the Promotion of Science KAKENHI, under Grant No. JP19H00651, JP19H01818, and JP19H05826 (K.O.). We acknowledge that this work was partly carried out in the Central Research Facilities at the Frederick Seitz Materials Research Laboratory, University of Illinois at Urbana-Champaign; we are grateful to the staff T. Shang, F. Rao, and M. Sardela (Central Research Facilities, Frederick Seitz Materials Research Laboratory, University of Illinois at Urbana-Champaign) for providing support during the sample preparation. Synchrotron SOLEIL is supported by the Centre National de la Recherche Scientifique (CNRS) and the Commissariat à l'Energie Atomique et aux Energies Alternatives (CEA), France. This work was also supported by a public grant by the French National Research Agency (ANR) as part of the "Investissements d'Avenir" (reference: ANR- 17-CE09-0016-05).

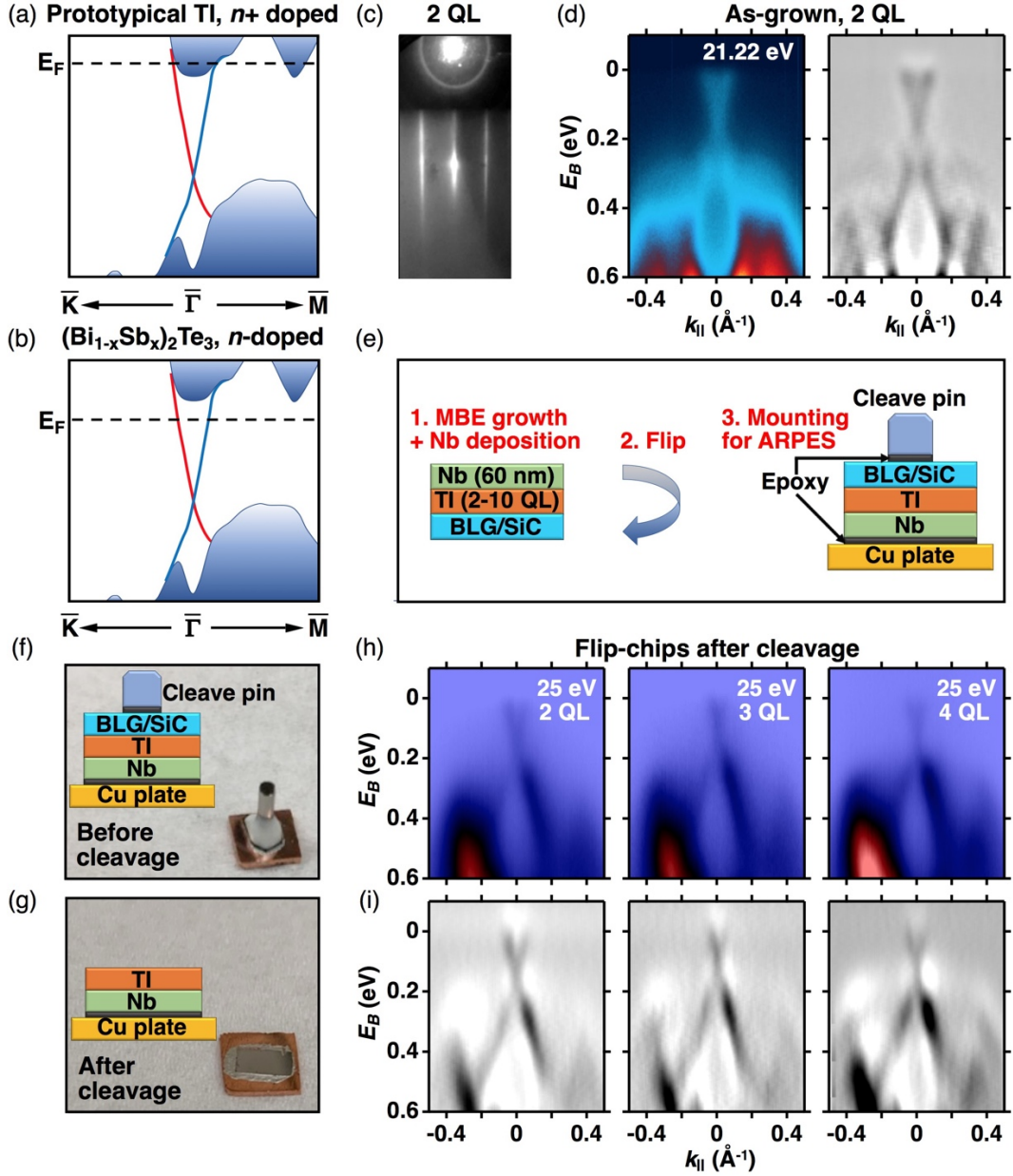

FIG. 1. Flip-chip method and characterizing lightly *n*-type TIs. (a)–(b) Schematic band structures of *n*+ doped prototypical TIs and lightly *n*-doped $(Bi_{1-x}Sb_x)_2Te_3$, respectively. Fermi levels are indicated with black dashed lines. (c) RHEED pattern of a 2 QL $(Bi_{1-x}Sb_x)_2Te_3$ film ($x = 0.62$) grown on BLG/SiC. (d) ARPES spectra (left panel) and corresponding second-derivative map (right panel) of 2 QL $(Bi_{1-x}Sb_x)_2Te_3$ taken along $\overline{\Gamma K}$ with 21.22-eV photons at 18 K. (e) Illustration of flip-chip technique for preparing TIs on superconducting Nb substrates. (f) Photo and diagram

of flip-chip sample structure before cleavage. (g) Similar as in (f) but after cleavage. (h) ARPES maps of 2–4 QL $(Bi_{1-x}Sb_x)_2Te_3$/Nb ($x = 0.62$) flip-chip samples taken with 25-eV photons at 80 K. (i) Corresponding second-derivative maps.

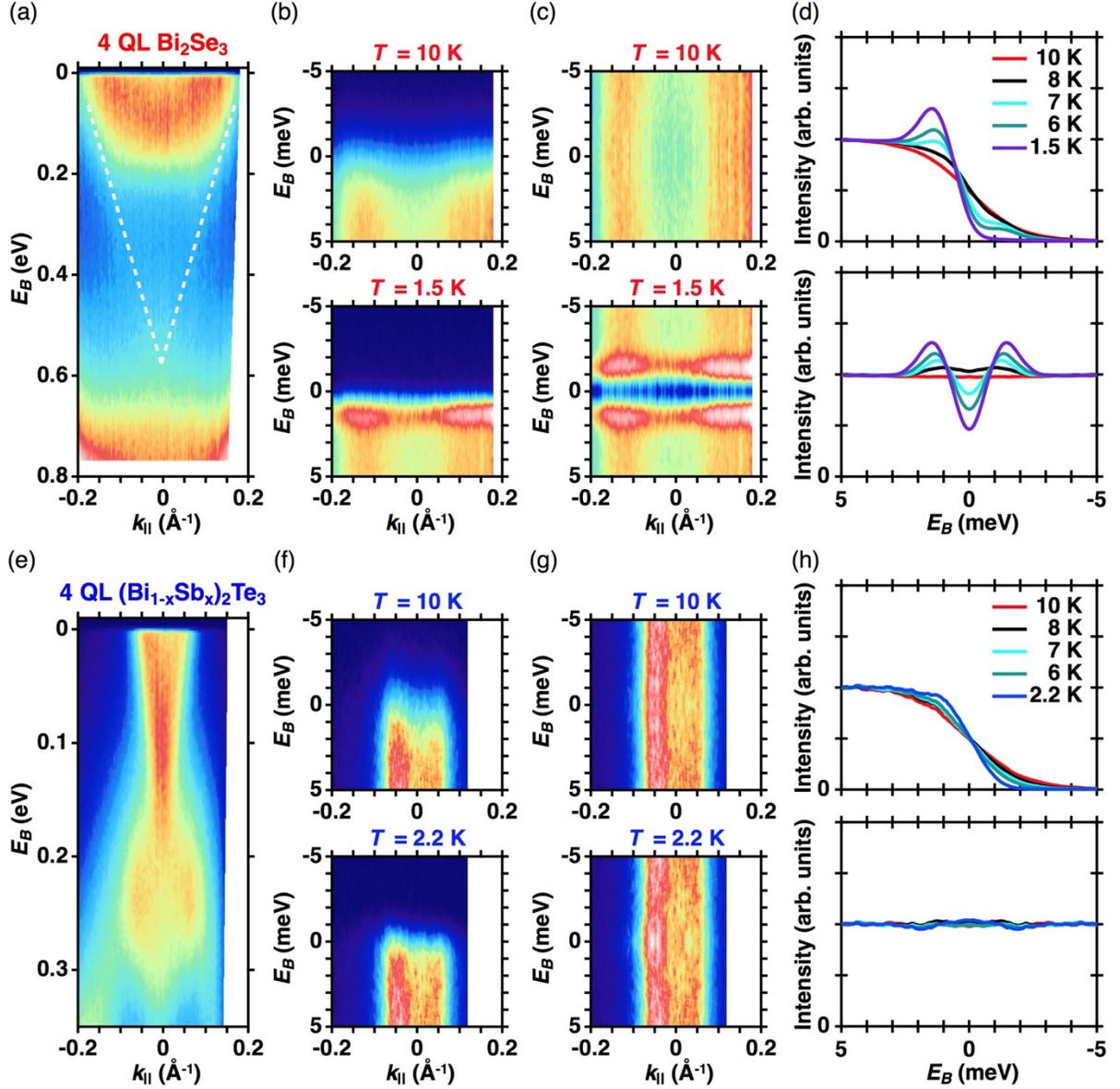

FIG. 2. Contrast between surface superconductivity in $n+$ doped TIs and $n$-type TIs. (a) ARPES map of 4 QL $Bi_2Se_3$/Nb taken with 6.994-eV photons at $T = 10$ K. The dashed white lines are guides to the eye that identify the Dirac cone. (b) Detailed ARPES spectra for binding energies near the Fermi level for 4 QL $Bi_2Se_3$/Nb at $T = 10$ K (top panel) and $T = 1.5$ K (bottom panel). (c) Corresponding symmetrized ARPES spectra, showing superconducting gaps and coherence peaks at all measured momenta at $T = 1.5$ K. (d) $k$-space integrated EDCs as a function of temperature (top panel) and corresponding symmetrized EDCs (bottom panel), with the $k$-space integration

taken over the Dirac cone (from -0.18 to 0.18 Å$^{-1}$). (e)–(h) Corresponding dataset for 4 QL (Bi$_{1-x}$Sb$_x$)$_2$Te$_3$/Nb for $x$ = 0.62, which reveals no superconductivity signatures; the $k$-space integration in (h) is taken over the Dirac cone (from -0.1 to 0.1 Å$^{-1}$).

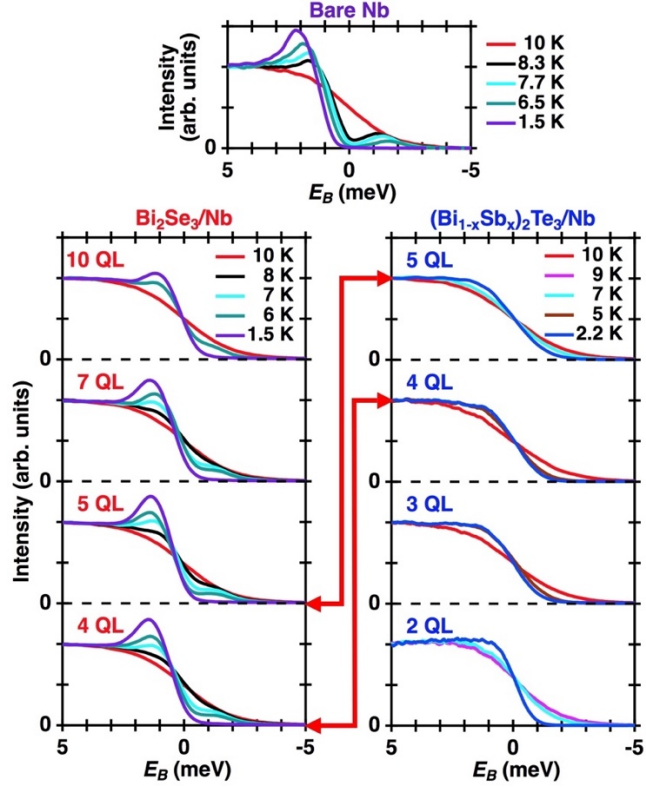

FIG. 3. *k*-space integrated EDCs as a function of temperature and thickness. Results are shown for a Nb reference (top panel), 4–10 QL $Bi_2Se_3$/Nb (bottom, left panel), and 2–5 QL $(Bi_{1-x}Sb_x)_2Te_3$/Nb for $x = 0.62$ (bottom, right panel). The *k*-space integrations are taken over -0.18 to 0.18 Å$^{-1}$ for $Bi_2Se_3$/Nb and -0.1 to 0.1 Å$^{-1}$ for $(Bi_{1-x}Sb_x)_2Te_3$/Nb. For comparisons between the measurements, the red arrows connect the spectra of 4 and 5 QL $Bi_2Se_3$/Nb to those of $(Bi_{1-x}Sb_x)_2Te_3$/Nb. Clear coherence peaks and leading edge shifts are visible in the Nb reference and $Bi_2Se_3$/Nb for all thicknesses, while none can be identified for all $(Bi_{1-x}Sb_x)_2Te_3$/Nb.

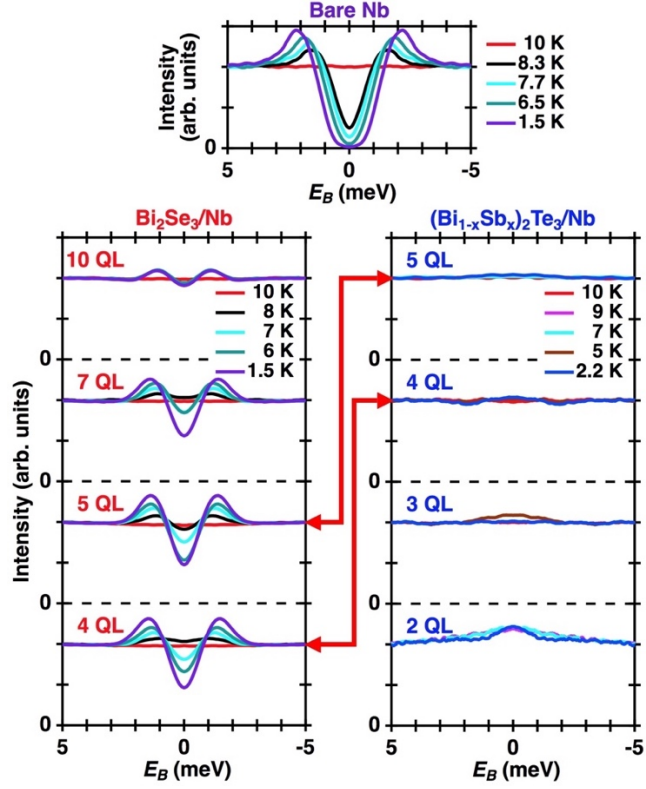

FIG. 4. Symmetrized *k*-space integrated EDCs as a function of temperature and film thickness. Results are summarized for a Nb reference (top panel), 4–10 QL $Bi_2Se_3$/Nb (bottom, left panel), and 2–5 QL $(Bi_{1-x}Sb_x)_2Te_3$/Nb for $x = 0.62$ (bottom, right panel). The *k*-space integrations are taken over -0.18 to 0.18 Å$^{-1}$ for $Bi_2Se_3$/Nb and -0.1 to 0.1 Å$^{-1}$ for $(Bi_{1-x}Sb_x)_2Te_3$/Nb. The red arrows are guides in the comparison between the datasets. As the temperature decreases from $T = 10$ K, a clear superconducting gap develops in the data of the Nb reference and all $Bi_2Se_3$/Nb, while none is evident for all $(Bi_{1-x}Sb_x)_2Te_3$/Nb samples.

# Supplemental Material for "Massive suppression of proximity pairing in topological (Bi$_{1-x}$Sb$_x$)$_2$Te$_3$ films on niobium"

Joseph A. Hlevyack, Sahand Najafzadeh, Meng-Kai Lin, Takahiro Hashimoto, Tsubaki Nagashima, Akihiro Tsuzuki, Akiko Fukushima, Cédric Bareille, Yang Bai, Peng Chen, Ro-Ya Liu, Yao Li, David Flötotto, José Avila, James N. Eckstein, Shik Shin, Kozo Okazaki, and T.-C. Chiang

**This files includes:**

**Supplementary text**

**A.** ARPES characterizations of as-grown films and flip-chip samples using a He discharge lamp

**B.** Synchrotron ARPES measurements at other photon energies of (Bi$_{1-x}$Sb$_x$)$_2$Te$_3$/Nb

**C.** Methods of Fermi-level determination and associated error analysis

**D.** Discussion about the choice of bulk insulating (Bi$_{1-x}$Sb$_x$)$_2$Te$_3$/Nb over (Bi$_{1-x}$Sb$_x$)$_2$Se$_3$/Nb

**E.** Supplementary references

**Figs. S1–S7**

**Tables SI–SII**

**Supplementary text**

**A. ARPES characterizations of as-grown films and flip-chip samples using a He discharge lamp**

Figure S1(a) shows an overview ARPES spectrum taken along $\overline{\Gamma K}$ of the bulk valence bands (left panel) of the as-grown 2 QL $(Bi_{1-x}Sb_x)_2Te_3$ film ($x = 0.62$) presented in Fig. 1(d) of the main text. A zoomed-in ARPES band map showing the gapped surface Dirac cone is also displayed [Fig. S1(a), right panel], and the corresponding second-derivative maps are pictured in Fig. S1(b). The clear gapped surface Dirac cone is again evidence of a smooth surface, and many sharp valence bands over a wide binding energy range (~0.3 to 4.5 eV) are evidence of a very well-ordered thin film. Generally, the results are consistent with what is expected for a telluride-based TI [1–5].

The cleavage of our flip-chip sample structure occurs exactly at the TI/substrate interface, resulting in a TI film on a bulk Nb substrate [see Figs. 1(e)–1(i) and also Ref. [6]]. Specifically, for TI films grown on BLG/SiC, little to no graphene layer is exfoliated off of the 6H-SiC(0001) substrate with the TI film during the cleavage [see Figs. S1(c) and S1(d)]. Fig. S1(c) shows an ARPES spectrum of an as-grown 3 QL $(Bi_{1-x}Sb_x)_2Te_3$ film for $x = 0.62$ (left panel) and its associated second-derivative map (right panel). For comparison, Fig. S1(d) displays a corresponding dataset for the same 3 QL film but now after the flip-chip sample preparation. Aside from intensity modulations introduced into the bands possibly caused by the substrate difference, the flip-chip spectrum in Fig. S1(d) still shows many sharp bulk valence bands that can all be identified with those in Fig. S1(c). Furthermore, no $\sigma$ band due to graphene bilayer at the zone center (at a binding energy of ~3 eV) is visible in the flip-chip map [7,8], suggesting that negligible graphene is cleaved away from the substrate's surface during the cleavage.

**B. Synchrotron ARPES measurements at other photon energies of $(Bi_{1-x}Sb_x)_2Te_3$/Nb**

We performed synchrotron measurements at other photon energies for 2–4 QL $(Bi_{1-x}Sb_x)_2Te_3$/Nb ($x = 0.62$), which verify that no bulk-like states cross the Fermi level in all samples measured. ARPES band maps taken at photon energies 21 eV and 30 eV along $\overline{\Gamma M}$ are displayed in Fig. S2 for 2 QL, Fig. S3 for 3 QL, and Fig. S4 for 4 QL $(Bi_{1-x}Sb_x)_2Te_3$/Nb. For comparison, the ARPES data taken with 25-eV photons as shown in the main text are reproduced in each figure. In all $(Bi_{1-x}Sb_x)_2Te_3$ films prepared on bulk Nb film, no conduction bands cross the Fermi level; here, the Fermi level lies in the bulk band gap, as expected for films at this composition [1].

**C. Methods of Fermi-level determination and associated error analysis**

This study requires exact quantification of small superconducting gaps in TIs/Nb and thus a precise determination of the Fermi level. An Au reference sample was used to calibrate the Fermi-level position via methods similar to those previously established [9,10]. The position of the Fermi level was determined at each detector channel by fitting the measured Au data with a Gaussian-broadened Fermi-Dirac distribution, assuming a cubic density of states (DOS). Fig. S5(a) shows the raw Au detector map (left panel) used for calibrating the 4 QL $(Bi_{1-x}Sb_x)_2Te_3$/Nb data and also a zoomed-in image of the same map overlaid with the fit, represented by a magenta curve, determined from the Fermi-level positions (right panel). This "curved" Fermi level was corrected by shifting the spectra at each detector channel, the result of which is shown in Fig. S5(b). Since the temperature-dependent deviation of the Au Fermi level from 0 K to 10 K is negligible (~-0.1 $\mu$eV, per the Sommerfeld expansion), this Au Fermi-level correction was applied to all TI/Nb spectra at all temperatures of interest (1.5 to 10 K); the Fermi-level stability was previously confirmed by repeatedly measuring the Au reference in intervals of ~1 hour, yielding a stability of about ±100 $\mu$eV [10]. Assuming a linear DOS in the Au Fermi-level fitting generally yielded a poor fit to the Au EDCs [Fig. S5(c)], while a fit using a cubic DOS was better [Fig. S5(d)].

All TI/Nb spectra taken near the Fermi level were subsequently $k$-space integrated over the Dirac cone (-0.1 to 0.1 Å$^{-1}$ for $(Bi_{1-x}Sb_x)_2Te_3$/Nb and -0.18 to 0.18 Å$^{-1}$ for $Bi_2Se_3$/Nb), yielding a set of temperature-dependent EDC curves [see Figs. S5(e) and S5(g) for the 3 and 4 QL $(Bi_{1-x}Sb_x)_2Te_3$/Nb datasets, respectively]. Since the superconducting transition temperature of Nb is $T_C$ = 9.3 K, the 10 K EDC of each TI/Nb system, when possible, was fitted using similar methods as the Au reference to double-check the position of the sample's Fermi level, which usually yielded a deviation of the sample's Fermi level away from the Au Fermi level [see the Fermi-level position $E_F$ determined from the 10 K fittings in Figs. S5(e) and S5(g)]. Since the work function of the sample and Au reference are different, the Fermi level of the sample's 10 K dataset was taken as the next reference point for both $Bi_2Se_3$/Nb and $(Bi_{1-x}Sb_x)_2Te_3$/Nb. However, because the superconducting gap is very small in our $(Bi_{1-x}Sb_x)_2Te_3$/Nb samples, the shift of the sample's Fermi level with temperature could be important and must be considered in the analysis to avoid errors in gap determination. These temperature-dependent shifts are analyzed using the Sommerfeld expansion, assuming linear Dirac bands for the topological surface states as a first approximation; this yielded, for instance, a Fermi-level shift from 2.2 K to 10 K of about -13 $\mu$eV. When possible, the Fermi level of $(Bi_{1-x}Sb_x)_2Te_3$/Nb was shifted from the Au-calibrated Fermi level according to these temperature-dependent deviations plus that deduced from the Au-calibrated 10 K data [see Figs. S5(f) and S5(h) for the final result], which tended to eliminate noticeable bumps or dips in the symmetrized EDC data [compare the bottom panels of Figs. S5(e) and S5(g) to the corresponding panels of Figs. S5(f) and S5(h), respectively]. Any remaining small bumps/wiggles in the symmetrized data are readily attributable to the stability of the Fermi level (~±100 $\mu$eV or so) during each measurement.

Incidentally, fits assuming higher-order DOS for the Au Fermi-level correction resulted in deviations of approximately ±5 $\mu$eV or less from the Fermi-level positions deduced using a cubic DOS, while those involving a quadratic or linear DOS yielded considerably larger deviations. See Tables SI and SII and also Fig. S6. Finally, Fig. S7 compares the results of our analysis procedures when assuming a linear DOS or a cubic DOS throughout the fitting procedures, underscoring the inadequacy of a linear DOS for the Fermi-level fittings.

**D. Discussion about the choice of bulk insulating $(Bi_{1-x}Sb_x)_2Te_3$/Nb over $(Bi_{1-x}Sb_x)_2Se_3$/Nb**

A few key considerations for the choice of TI thin-film materials had to be examined when designing the experiments for this study. The ideal scenario would be a comparison of $(Bi_{1-x}Sb_x)_2Se_3$/Nb with $Bi_2Se_3$/Nb, the latter being the first flip-chip system measured by ultrahigh-resolution ARPES [6]. Nevertheless, fabricating bulk insulating, topological $(Bi_{1-x}Sb_x)_2Se_3$ films is problematic, because the phase diagram is complicated by a topological phase transition from a TI to trivial insulator phase and a concomitant structural phase transition from a rhombohedral to orthorhombic lattice as the alloy ratio $x$ increases from $x = 0$ to 1 [11,12]. MBE-grown $(Bi_{1-x}Sb_x)_2Se_3$ films also usually exhibit mixed orthorhombic/rhombohedral phases over a wide composition range [11]. Taken together, these factors would obscure the mechanism for the superconducting proximity effect of fundamental interest in this work. Consequently, we chose the well-known, bulk insulating telluride-based TIs $(Bi_{1-x}Sb_x)_2Te_3$, wherein the Fermi level can be readily tuned by varying the alloy ratio *x* while avoiding the complications discussed for $(Bi_{1-x}Sb_x)_2Se_3$ [1–4]. Moreover, replacing Se with Te in these TI materials does not alter our conclusions presented in the main text, since the band structures of the selenide- and telluride-based TIs, including the topological surface states, are quite similar, provided that the lattice structures are the same [1–6,11,12].

**E. Supplementary references**

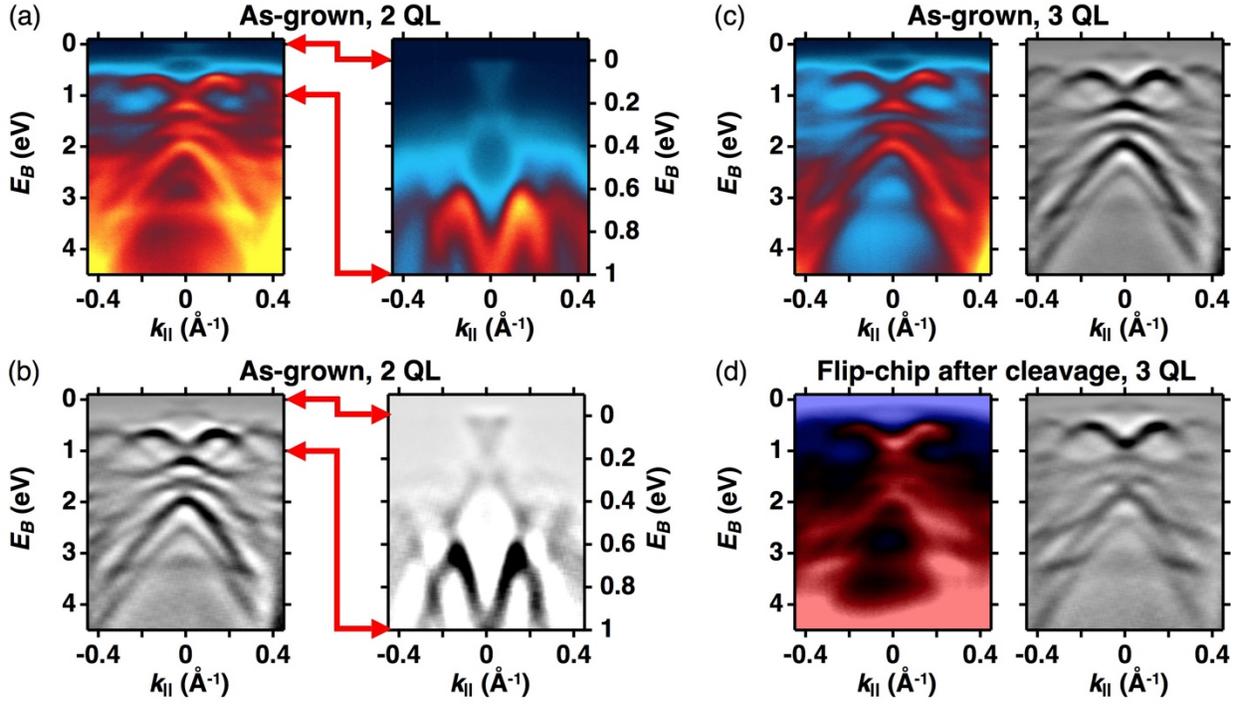

FIG. S1. ARPES spectra of as-grown and flip-chip $(Bi_{1-x}Sb_x)_2Te_3$ films taken with a He lamp. (a) ARPES map of an as-grown 2 QL $(Bi_{1-x}Sb_x)_2Te_3$ film ($x = 0.62$), showing bulk valence bands (left panel) and the gapped surface Dirac cone (right panel); the data were taken along $\overline{\Gamma K}$ using 21.22-eV photons at 18 K. (b) Corresponding second-derivative maps. (c) ARPES band map of an as-grown 3 QL $(Bi_{1-x}Sb_x)_2Te_3$ film ($x = 0.62$), measured at 18 K with 21.22-eV photons (left panel) and associated second-derivative map (right panel). (d) Same as in (c) for the same film but for a flip-chip sample.

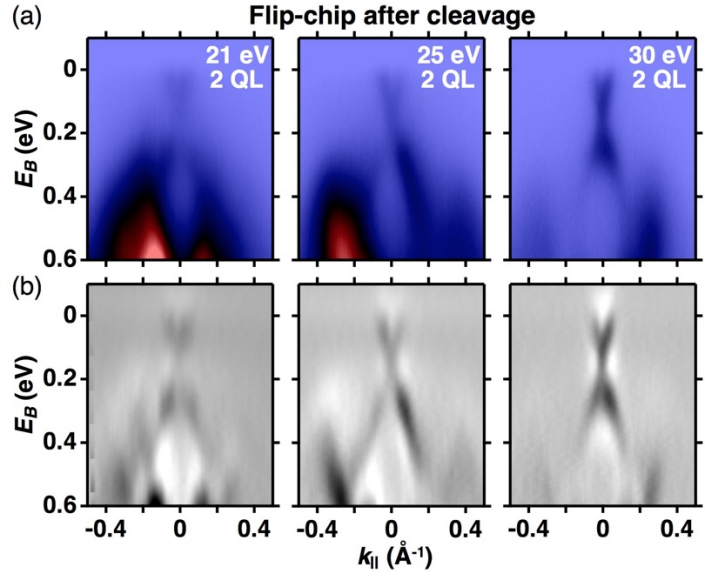

FIG. S2. Photon energy dependence of 2 QL $(Bi_{1-x}Sb_x)_2Te_3$/Nb. (a) Synchrotron ARPES measurements of 2 QL $(Bi_{1-x}Sb_x)_2Te_3$/Nb ($x = 0.62$) taken along $\overline{\Gamma M}$ with 21-eV (left), 25-eV (middle), and 30-eV (right) photons at 80 K. (b) Corresponding second-derivative maps.

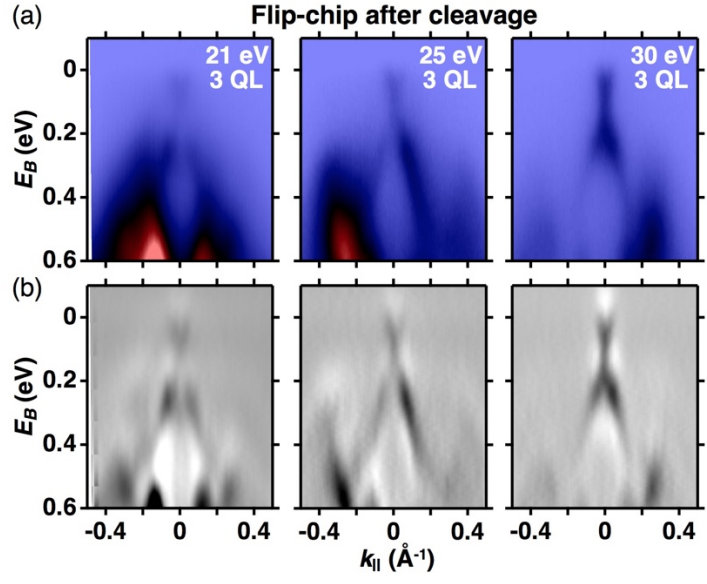

FIG. S3. Photon energy dependence of 3 QL $(Bi_{1-x}Sb_x)_2Te_3$/Nb. (a) Synchrotron ARPES spectra of 3 QL $(Bi_{1-x}Sb_x)_2Te_3$/Nb ($x = 0.62$) measured along $\overline{\Gamma M}$ using 21-eV (left), 25-eV (middle), and 30-eV (right) photons at a sample temperature of 80 K. (b) Corresponding second-derivative maps.

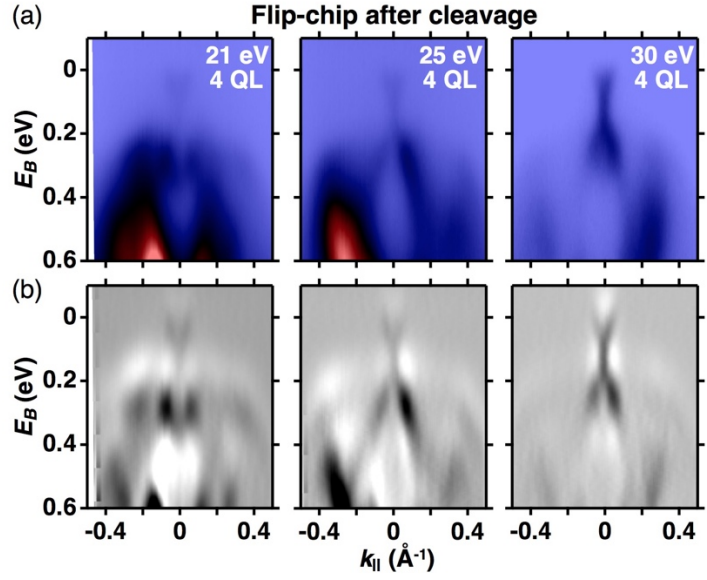

FIG. S4. Photon energy dependence of 4 QL $(Bi_{1-x}Sb_x)_2Te_3$/Nb. (a) Synchrotron ARPES band maps of 4 QL $(Bi_{1-x}Sb_x)_2Te_3$/Nb ($x = 0.62$); the data was taken along $\overline{\Gamma M}$ with 21-eV (left), 25-eV (middle), and 30-eV (right) photons at 80 K. (b) Corresponding second-derivative maps.

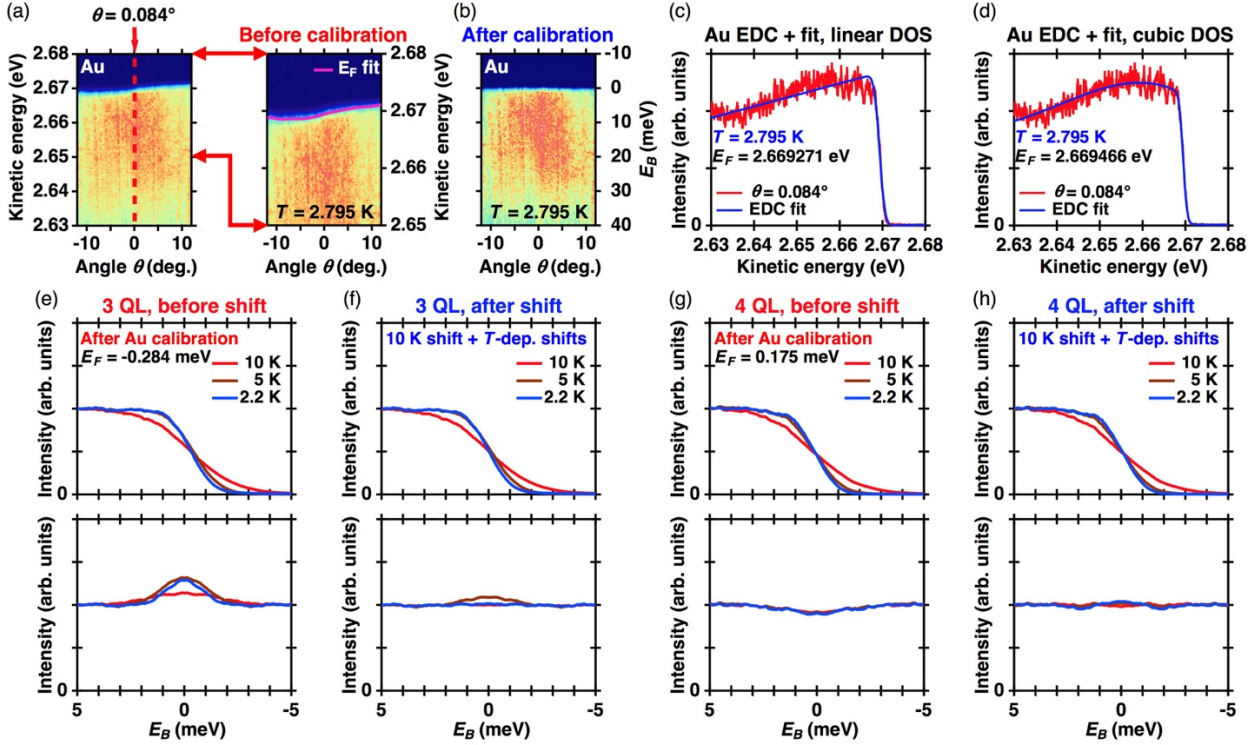

FIG. S5. Overview of analysis procedures for determining the Fermi-level position. (a) Measured ARPES map (photon energy of 6.994 eV) from a Au reference sample. The left panel is the raw data, and the right shows a zoomed-in image overlaid with the deduced Fermi-level fit, assuming a cubic DOS (magenta curve); the red dashed line in the left panel defines an EDC slice used for comparing fits generated with different DOS [see (c) and (d)]. (b) Au ARPES map from (a) with the Fermi-level correction to line up the Fermi level for the different emission angles. (c) Au EDC from (a) fitted using a linear DOS. (d) Same Au EDC as in (c) but fitted assuming a cubic DOS. (e)–(f) $k$-space integrated EDCs (top panel) and associated symmetrized EDCs (bottom panel) as a function of temperature for 3 QL $(Bi_{1-x}Sb_x)_2Te_3$/Nb ($x = 0.62$), before (e) and after (f) the Fermi-level shifts; the $k$-space integration is from -0.1 to 0.1 Å$^{-1}$. The $E_F$ value in (e) is found from fitting the 10 K data (rounded here to three significant figures for simplicity). (g)–(h) Same as (e)–(f) but for 4 QL $(Bi_{1-x}Sb_x)_2Te_3$/Nb ($x = 0.62$).

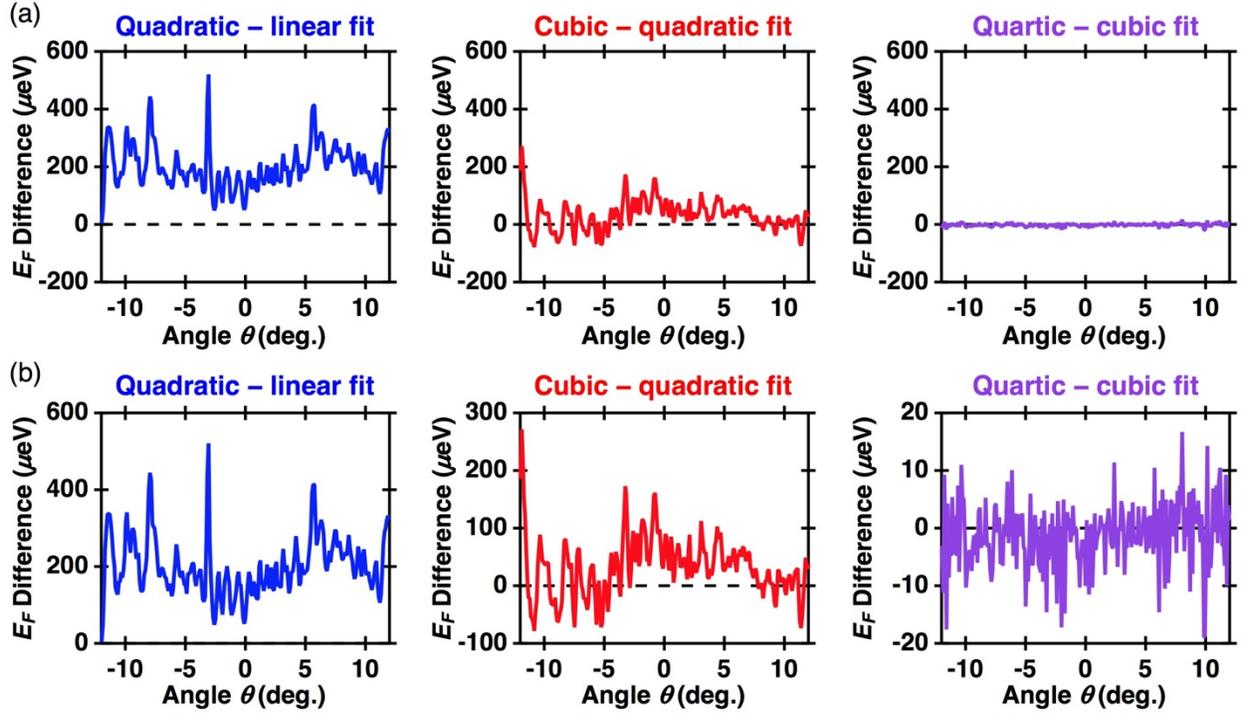

FIG. S6. Differences between the Fermi-level fit positions using other DOS. (a) Differences between the fitted Fermi-level positions as a function of detector angle for those fits using a quadratic and a linear DOS (left panel), a cubic and a quadratic DOS (middle panel), and a quartic and a cubic DOS (right panel); this dataset is for the Au reference used to calibrate the Fermi level of 4 QL $(Bi_{1-x}Sb_x)_2Te_3$/Nb. (b) Same as in (a) but zoomed in to highlight the decreasing differences with increasing DOS polynomial order; the difference between the fitted Fermi-level positions assuming quartic and cubic DOS is nearly negligible.

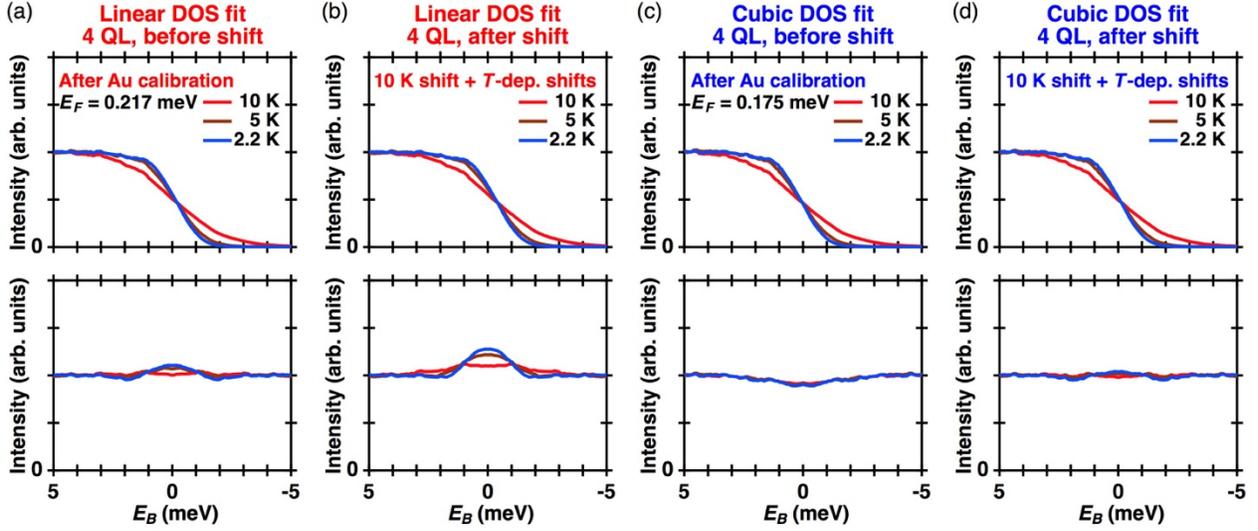

FIG. S7. Large error in the Fermi-level fitting when assuming a linear DOS. (a)–(b) $k$-space integrated EDCs (top panel) and corresponding symmetrized ARPES data (bottom panel) of 4 QL $(Bi_{1-x}Sb_x)_2Te_3$/Nb ($x = 0.62$) as a function of temperature, generated assuming a linear DOS in the Fermi-level fitting both before (a) and after (b) the Fermi-level shifts described in the Supplementary text; the $k$-space integration is from -0.1 to 0.1 Å$^{-1}$, and for visual clarity in (a), the Fermi level $E_F$ deduced from fitting the 10 K data is rounded to three significant figures. In the symmetrized data corresponding to fits with a linear DOS [bottom panel of (b)], an erroneous peak in the data arises very prominently at the Fermi level. (c)–(d) Same as in (a)–(b), respectively, but assuming a cubic DOS during the Fermi-level fitting. The erroneous peak in (b) becomes very much suppressed. The remaining noise-like features are well within the experimental uncertainty of the measurement.

| | Au fit parameters, $\theta = 0.084°$, $T = 2.795$ K | | | |
|---|---|---|---|---|
| | Linear DOS | Quadratic DOS | Cubic DOS | Quartic DOS |
| $E_F$ | 2.669271 eV | 2.669376 eV | 2.669466 eV | 2.669467 eV |
| $\Delta E$ | 2.170 meV | 1.792 meV | 1.456 meV | 1.462 meV |

TABLE SI. Comparison between the Fermi-level fitting parameters with different DOS. These fitting parameters are for a particular angular slice of the Au reference spectra, presented here as an example to illustrate the differences from different fitting assumptions. Here, $E_F$ and $\Delta E$ are the Fermi level and the system energy resolution deduced from the fitting procedure, respectively. The Au reference spectra which generated this table was utilized during the analysis of the 4 QL $(Bi_{1-x}Sb_x)_2Te_3$/Nb data appearing in the main text.

| | Au fit parameter differences | | |
|---|---|---|---|
| | Linear → quadratic | Quadratic → cubic | Cubic → quartic |
| Avg. change in $E_F$ | 195 $\mu$eV | 32 $\mu$eV | -1 $\mu$eV |
| Avg. change in $\Delta E$ | -728 $\mu$eV | -123 $\mu$eV | 6 $\mu$eV |

TABLE SII. Differences between fit parameters of different Au Fermi-level fittings. Here, $E_F$ and $\Delta E$ are the Fermi-level positions and the total system energy resolution deduced from the fitting procedures, respectively; the differences are taken between fit parameters obtained assuming a different DOS. The Au reference spectra which generated this table was used for the Fermi-level corrections in the 4 QL $(Bi_{1-x}Sb_x)_2Te_3$/Nb data in the main text.